\documentstyle[aps,prl,epsf,twocolumn,floats]{revtex}

\begin{document}
\draft

%%2col
\twocolumn[\hsize\textwidth\columnwidth\hsize\csname @twocolumnfalse\endcsname
%% start of wide text
%%2col

\title{Mixed-valent regime of the two-channel Anderson impurity
       as a model for UBe$_{13}$}
\bigskip
\author{Avraham Schiller,$^1$
        Frithjof B. Anders,$^{2,3}$
        and Daniel L. Cox$^3$}
\address{$1$ Department of Physics, The Ohio State University,
             Columbus, OH 43210-1106\\
         $2$ Institut f\"ur Festk\"orperphysik, Technical
             University Darmstadt, 64289 Darmstadt, Germany\\
         $3$ Department of Physics, University of California,
             Davis, California 95616}
\date{\today}
\maketitle

\begin{abstract}
We investigate the mixed-valent regime of a two-configuration Anderson
impurity model for uranium ions, with separate quadrupolar and
magnetic doublets. With a new Monte Carlo approach and the
non-crossing approximation we find: (i) A non-Fermi-liquid fixed point
with two-channel Kondo model critical behavior; 
(ii) Distinct energy scales for screening the low-lying and
excited doublets; (iii) A semi-quantitative explanation of 
magnetic-susceptibility data for U$_{1-x}$Th$_x$Be$_{13}$
assuming 60-70\% quadrupolar doublet ground-state weight,
supporting the quadrupolar-Kondo interpretation. 
\end{abstract}

\bigskip
%% 71.27.+a -- Strongly correlated electron systems; heavy fermions.
%% 75.30.Mb -- Valence fluctuation, Kondo lattice, and heavy-fermion phenomena.
%% 75.20.Hr -- Local moment in compounds and alloys; Kondo effect, valence
%%             fluctuations, heavy fermions.

%%2col
%% end of wide text
]
\narrowtext
%%2col

% Paragraph 1: General introduction.
Since the 1950's, Landau's Fermi-liquid theory has shaped our understanding
of the metallic state. Based on the notion of a one-to-one mapping
between the low-lying excitations of the interacting system and that
of the noninteracting electron gas, the theory provides a remarkably
robust scenario for the low-temperature properties of interacting
electron systems. It is against this outstanding success that a growing
class of $f$-shell materials --- predominantly Ce- and U-based alloys
--- received considerable attention in recent years. Characterized by
a logarithmically divergent linear coefficient of specific heat
and anomalous temperature dependences of the resistivity and
susceptibility, these materials appear to depart from the conventional
Fermi-liquid scenario~\cite{Maple95}, thus challenging our understanding
of metallic behavior.

%Paragraph 2: Overview of results.
In this paper, we present results on the two-channel Anderson impurity
model in the mixed-valent regime, motivated by the unusual nonlinear
susceptibility data of the non-Fermi-liquid (NFL) alloy system
U$_{1-x}$Th$_x$Be$_{13}$. The restricted Hilbert space of our model, a
ground quadrupolar (non-Kramers) doublet in the $5f^2$ configuration
and a ground magnetic (Kramers) doublet in the $5f^3$ configuration,
renders it intractable to study by conventional Monte Carlo methods.  
We have developed a new Monte Carlo method based upon the mapping
onto a Coulomb gas. We use this method to calibrate non-crossing
approximation (NCA) results, which can then be extended to more
extreme parameter regimes. We find that the model displays NFL physics
characteristic of the two-channel Kondo model, even at the extreme
mixed-valent limit when the two charge configurations are degenerate.
We also find that two energy scales appear in the screening process
away from the configurational degeneracy point.
Using this model we are able to semi-quantitatively explain the linear
and nonlinear susceptibility data by assuming 60-70\% ground-state weight
for the quadrupolar doublet, with the Th doping inducing a higher $5f^2$
count, consistent with expectations from lattice constant data. We do not,
however, explain the small energy scale for UBe$_{13}$, and anticipate
that dynamical inclusion of excited crystal field (CEF) levels in the
$5f^2$ configuration may remedy this problem.

%Paragraph 3: The mixed-valent scenario.
A proposed scenario for the NFL physics of UBe$_{13}$
involves the screening of uranium quadrupole moments in the $5f^2$
configuration by conduction orbital motion~\cite{Cox87}. This quadrupolar
Kondo effect can explain the enhanced specific heat and other data
in this compound. In principle, though, the screened moment in the
scenario of Ref.~\onlinecite{Cox87} can either be magnetic or quadrupolar,
depending on which ionic charge configuration (U$^{3+}$ or U$^{4+}$)
is lower in energy. Which case is realized in UBe$_{13}$ is still
unresolved. While the nonlinear magnetic susceptibility~\cite{Ramirez94}
is suggestive of a magnetic trivalent state,
the rather weak temperature dependence of the linear susceptibility
is more consistent with a nonmagnetic tetravalent state. Recently,
following experiments on the nonlinear susceptibility of
U$_{0.9}$Th$_{0.1}$Be$_{13}$, Aliev {\em et al.}~\cite{Aliev95} suggested
a different physical regime, in which strong quantum fluctuations
drive the uranium ions to a mixed-valent state.

%Paragraph 4: The model.
Here we explore the possibility of a mixed-valent state in the framework
of the two-channel Anderson model~\cite{CZ97}, which consists of $\Gamma_8$
conduction electrons carrying both spin ($\sigma = \uparrow,\downarrow$)
and quadrupolar ($\alpha = \pm$) quantum labels, hybridizing via a matrix
element $V$ with a local uranium ion. The latter is modeled by a $\Gamma_3$
quadrupolar doublet in the $5f^2$ configuration and a $\Gamma_6$ magnetic
doublet in the $5f^3$ configuration, separated in energy by
$\epsilon_f = E(5f^2) - E(5f^3)$. The corresponding Hamiltonian reads
\begin{eqnarray}
{\cal H} &=& \sum_{k\alpha\sigma}
             \epsilon_k c^{\dagger}_{k\alpha\sigma} c_{k\alpha\sigma}
          +  \epsilon_f\sum_{\alpha}
             \left |5f^2, \alpha\right>\left<5f^2, \alpha \right|
\label{Hamiltonian} \\
         &+& V \sum_{k\alpha\sigma} \left\{
             c^{\dagger}_{k\alpha\sigma} \left |5f^2,-\alpha \right>
             \left < 5f^3 , \sigma \right| + {\rm h.c.} \right\} ,
\nonumber
\end{eqnarray}
where $c^{\dagger}_{k\alpha\sigma}$ creates a $\Gamma_8$ conduction
electron with spin $\sigma$ and quadrupolar moment $\alpha$. This model
may be applied to U$_{1-x}$Th$_x$Be$_{13}$ for temperatures above about
half the Kondo scale, for which coherent lattice effects are small.  

%Paragraph 5: Technical challenge.
In the integer-valence limit, $\Gamma \equiv \pi \rho V^2 \ll
|\epsilon_f|$, ($\rho$ being the conduction-electron density of
states at the Fermi level) Eq.~(\ref{Hamiltonian}) reduces
to the two-channel Kondo Hamiltonian~\cite{NB80}, which is exactly
solvable by a number of methods: Bethe ansatz~\cite{Bethe_ansatz},
conformal field theory~\cite{CFT}, and Abelian bosonization~\cite{EK92}.
However, none of these solutions extend to the Hamiltonian of
Eq.~(\ref{Hamiltonian}), where charge fluctuations are present.
This model is also intractable to conventional determinantal Quantum
Monte Carlo. 

%Paragraph 6: Our approach.
To overcome the difficulties associated with the Hamiltonian of
Eq.~(\ref{Hamiltonian}) we devised a new approach, based on (i)
mapping the corresponding partition function onto a classical
one-dimensional Coulomb gas, and (ii) sampling the latter gas using
Monte Carlo techniques. This approach allows an accurate calculation of
the low-temperature thermodynamics of the model deep into the
mixed-valent regime, all the way from weak to strong coupling. It
also enables the computation of the nonlinear susceptibility, which,
due to the high-order correlation function involved, is typically
inaccessible to determinantal Quantum Monte Carlo. Below we outline
our approach, starting with the mapping onto a Coulomb gas.

%Paragraph 7: Kink-gas representation --- introduction.
The formal connection between the Kondo effect and the
statistical-mechanical problem of a one-dimensional Coulomb gas
was first recognized for the Kondo Hamiltonian~\cite{AY69},
and later extended to the one-channel Anderson model~\cite{Haldane78,SK93}.
Here we generalize the mapping to the Hamiltonian of
Eq.~(\ref{Hamiltonian}). The basic idea is to expand the partition
function in powers of $V$, expressing it as a sum over all possible
histories of the impurity. A history is a sequence of hopping events,
or kinks, which overall preserve the impurity state and the
occupation of each conduction-electron branch. Each history is
represented by a sequence of impurity states,
$\{ \gamma_0, \ldots ,\gamma_n \}$, and a sequence of imaginary times,
$\{ \tau_1, \ldots ,\tau_n \}$, corresponding to the instances at
which hopping events take place. Tracing over the conduction-electron
degrees of freedom generates a long-range ``Coulomb'' interaction
between the kinks within a given history. These couple through the
four-component charges $\vec{\epsilon}_i = (\delta N_{\uparrow +},
\delta N_{\downarrow +}, \delta N_{\uparrow -}, \delta N_{\downarrow -})$,
where $\delta N_{\sigma\alpha}$ is the change in occupancy of the
$\sigma$, $\alpha$ conduction-electron branch due to the $i$-th hopping
event (i.e., the transition from state $\gamma_{i-1}$ to state $\gamma_i$).
Denoting the impurity-free partition function by $Z_0$, one has
\begin{eqnarray}
&&\frac{Z}{Z_0} = \sum_{n = 0}^{\infty}
    \left(\frac{\Gamma}{\pi D}\right)^{n/2}\!\!\!\!
    \sum_{\gamma_n = \gamma_0, \ldots, \gamma_{n-1}}\!\!
    \delta_{\Sigma_i \vec{\epsilon}_i, 0}
    \int_{0}^{\beta}\! \frac{d\tau_n}{\tau_c}
    \ldots \int_{0}^{\tau_2}\!\frac{d\tau_1}{\tau_c}
\nonumber\\
&&\times \exp
    \left [ \sum_{i > j = 1}^n F(\tau_i - \tau_j) \vec{\epsilon}_i\cdot
    \vec{\epsilon}_j
    - \sum_{i = 0}^{n} (\tau_{i+1} - \tau_i) E_{\gamma_i} \right] ,
\label{kink-gas}
\end{eqnarray}
\noindent where $E_{\gamma}$ is the bare energy of the $\gamma$ impurity
state (zero for $5f^3$ and $\epsilon_f$ for $5f^2$);
$F(\tau)$ is the interaction strength between kinks with time separation
$\tau$; and $\tau_0$ and $\tau_{n+1}$ are equal to zero and $\beta$,
respectively. At zero temperature, $F(\tau)$ reduces to
$\ln (1 + |\tau|/\tau_c)$, where $D = 1/\tau_c$ plays the role of a
bandwidth. At $T > 0$, the logarithm is replaced by a more complicated
expression. The neutrality condition
$\delta_{\Sigma_i \vec{\epsilon}_i, 0}$ selects only those histories that
overall preserve the occupation of each conduction branch.

%Paragraph 8: No sign problem.
The impurity contribution to thermodynamic quantities can be computed
directly from the Coulomb-gas representation of Eq.~(\ref{kink-gas})
using Monte Carlo. This formulation has the crucial advantage of being
free of any sign problem, as the different terms in Eq.~(\ref{kink-gas})
are all positive. Below we present our results for the mixed-valent regime.

\begin{figure}
\centerline{
\vbox{\epsfxsize=90mm \epsfbox {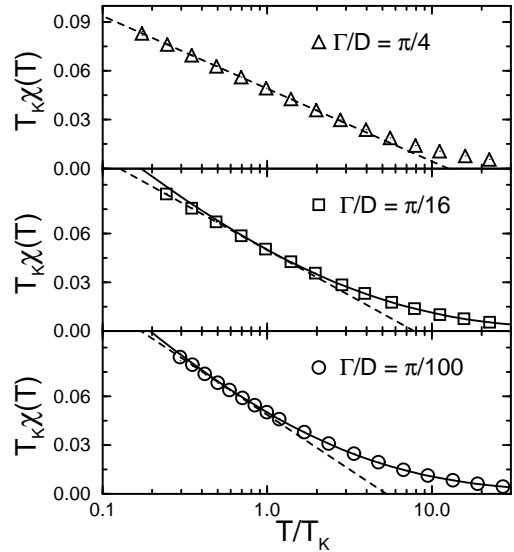}}
}
\caption{The impurity susceptibility, $\chi(T)$, for $\epsilon_f = 0$
and different $\Gamma/D$. $\chi(T)$ is defined as the response to a
field that couples linearly to one of the impurity moments
(for $\epsilon_f = 0$, the magnetic and quadrupolar susceptibilities
are identical). Error bars are smaller than the symbols used. For $T < T_K$,
$T_K$ the Kondo scale, $\chi(T)$ has the logarithmic temperature
dependence $\chi(T) \sim (a/T_K) \log(T_K/T)$, with $a$ extracted from a
logarithmic fit (dashed lines). In going from large to small $\Gamma/D$,
$a$ takes the values $0.019, 0.024$, and $0.029$, while $T_K/D$ equals
$0.09, 0.032$, and $0.00662$, respectively. The solid lines in the
middle and lower graphs are the results of
the NCA with $\Gamma/D = \pi/23$ and $\Gamma/D = \pi/144$, respectively
(see text). There is excellent agreement between the NCA and the Monte
Carlo down to $T/T_K \approx 0.7$, at which point the NCA curves
cross over to a logarithmic temperature dependence with $a = 0.037$ and
$0.042$ for $\Gamma/D = \pi/23$ and $\Gamma/D = \pi/144$, respectively.}
\end{figure}

%Paragraph 9: Degeneracy point --- chi(T).
We begin our discussion with the limit of strong valence fluctuations,
$|\epsilon_f| \ll \Gamma$, represented by $\epsilon_f = 0$. In
Fig.~1 we have plotted the impurity susceptibility, $\chi(T)$,
in response to a field that couples linearly to one of the impurity moments
--- either the magnetic moment in the case of a magnetic field, or the
quadrupolar moment in the case of an electric or strain field (the two
susceptibilities are identical for $\epsilon_f = 0$). In a Fermi liquid,
$\chi(T)$ saturates at a constant as $T \to 0$. Here it diverges
logarithmically with decreasing temperature, consistent with
the NFL behavior of the two-channel Kondo model~\cite{Bethe_ansatz}.
The logarithmic temperature dependence extends
from weak coupling ($\Gamma/D\approx 0.03$ in Fig.~1) to strong
coupling ($\Gamma/D \approx 0.8$). Thus, the same critical behavior of
the integer-valent limit persists into the mixed-valent regime.

%Paragraph 10: Definition of the Kondo scale.
The crossover to logarithmic temperature dependence in Fig.~1 is
associated with a characteristic energy scale or Kondo temperature, $T_K$,
defined as the temperature at which the effective moment per unit
occupancy, $\mu_{eff}(T) \equiv T\chi_i(T)/n_i(T)$, is 60\% screened.
Here $n_i(T)$ and $\chi_i(T)$ are the occupancy and susceptibility
of the corresponding $5f^i$ doublet
($i=2,3$; we use $\mu_B g_i = 1$, except in Fig.~3). Note that the
$f$-electron count is given by $3-n_2(T)$, and that $n_i(T)$ is fixed
at one half for $\epsilon_f = 0$. At high temperature, $\mu_{eff}(T)$
reduces to the free-moment value of one quarter. Hence $T_K$ is defined
by $\mu_{eff}(T_K) = 0.1$. For the cases shown in Fig.~1,
this definition gives good agreement (within 25\%) with a slope of
$1/(20 T_K)$ for the logarithmic component of $\chi_i(T)/n_i(T)$,
as is characteristic of the two-channel Kondo effect~\cite{SS89}. 

%Paragraph 11: Lack of small Kondo scale.
Figure~2 depicts the Kondo temperature $T_K$ as a function
of $\Gamma$, for $\epsilon_f = 0$. Rather than varying exponentially with 
$1/\Gamma$, over the range $0.01 < \Gamma/D < 0.2$ we find the 
power-law dependence $T_K \sim \Gamma^x$, with $x \approx 0.9$.
This behavior is very close to the linear dependence expected from
simple renormalization-group (and analytic NCA) arguments. However, we
find  $T_K/\Gamma \approx 0.2$, which is not expected from the
renormalization-group and NCA arguments.  
For a representative value $\Gamma = 0.3{\rm eV}$, this
ratio gives a Kondo temperature of $T_K \sim 600{\rm K}$, i.e.,
sixty-fold larger than the $10{\rm K}$ seen in UBe$_{13}$. Hence,
contrary to previous claims for an uranium ion with full spherical
symmetry~\cite{spherically_symmetric_ion}, our model
does not support a small energy scale in the
limit of strong valence fluctuations.

\begin{figure}
\centerline{
\vbox{\epsfxsize=75mm \epsfbox {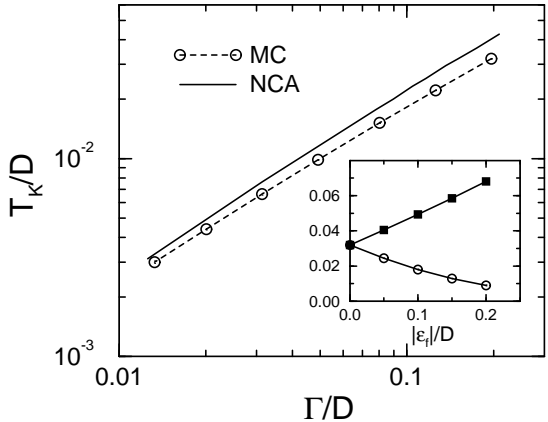}}
}
\caption{$T_K$ versus $\Gamma$, for $\epsilon_f = 0$. Open circles are
results of the Monte Carlo; full line was obtained using the NCA. Error
bars for the Monte Carlo data are smaller than the symbols used. Inset: The
screening temperatures $T_K/D$ (open circles) and $T_{ex}/D$ (filled squares)
as a function of $|\epsilon_f|/D$, for fixed $\Gamma/D = \pi/16$ (results
obtained using Monte Carlo).}
\end{figure}

%Paragraph 12: Comparison to NCA.
Figure~2 also presents a comparison between the Monte Carlo approach
and the NCA~\cite{CZ97}. Within the NCA, the low-energy physics is
exclusively determined by the ratio of the spin degeneracy to the
number of independent conduction-electron channels~\cite{CZ97}. Thus,
irrespective of the inter-configurational energy $\epsilon_f$, the
NCA yields the critical behavior of the two-channel Kondo model for
the Hamiltonian of Eq.~(\ref{Hamiltonian}), in agreement with the
Monte Carlo result. In Fig.~2, the NCA curve for $T_K$ is slightly
shifted (on a log-log scale) with respect to the Monte Carlo one.
This shift may be accounted for by rescaling $\Gamma^{NCA}$ relative
to $\Gamma^{MC}$, which we attribute in part to the different high-energy
cut-off schemes used in the two formulations. By matching the NCA and
the Monte Carlo occupation numbers away from $\epsilon_f = 0$, we
extracted the rescaling factor $\Gamma^{MC}/\Gamma^{NCA} \approx 1.44$.
Using this rescaling, one obtains good quantitative agreement between
the Monte Carlo and the NCA throughout the mixed-valent regime, as is
exemplified in Fig.~1 for the scaled susceptibility, $T_K \chi(T)$.

%Paragraph 13: Introducing the two screening temperatures for e_f \neq 0.
In going to $\epsilon_f \neq 0$, the spin and the quadrupolar moments are
no longer screened at the same temperature. Specifically, the moment
associated with the excited doublet is quenched first at a characteristic
temperature $T_{ex}$, followed by the moment associated with the low-lying
doublet which is screened at $T_K < T_{ex}$. Physically, each
screening temperature represents a different crossover. Screening of the
excited doublet sets in as the occupancy of that doublet crosses over
from a high-temperature free-ion (Boltzmann) form to one that is governed
by valence fluctuations. On the other hand, $T_K$ marks the crossover to
strong coupling and the onset of NFL behavior.

%Paragraph 14: Inset of Fig. 2.
In the inset of Fig.~2 we have plotted $T_{ex}$ and $T_K$
as a function of $|\epsilon_f|$, for $\Gamma/D = \pi/16$.
The ground-state occupancy of the $5f^2$ doublet ranges in this
plot from $n_2 = 0.29$ for $\epsilon_f = 0.2$ to $n_2 = 0.71$
for $\epsilon_f = -0.2$. Note that $T_{ex}$ and $T_K$ interchange
meanings, depending on the sign of $\epsilon_f$: for $\epsilon_f > 0$,
$T_{ex}$ and $T_K$ are the screening temperatures for the quadrupolar
and the magnetic moment, respectively; for $\epsilon_f < 0$, the roles
are reversed.

%Paragraph 15: Discussion of the two screening temperatures.
As seen in Fig.~2, $T_{ex}$ grows essentially linearly with $|\epsilon_f|$.
This stems from the fact that the free-ion occupancy of the excited
doublet decays exponentially with $|\epsilon_f|/T$. Consequently, the
crossover to an occupancy driven by charge fluctuations is exponentially
sensitive to $|\epsilon_f|/T_{ex}$. Contrary to $T_{ex}$, $T_K$ decreases
with increasing $|\epsilon_f|$, becoming exponentially small
for $|\epsilon_f| \gg \Gamma$. Extracting the ratio $T_K/\Gamma$ for
$|\epsilon_f| = 0.2$ and using the value $\Gamma = 0.3{\rm eV}$ one
obtains $T_K \sim 150{\rm K}$, which is still an order of magnitude
too large to explain the $10{\rm K}$ observed in UBe$_{13}$.  

%Paragraph 16: Details of the calculation of the nonlinear susceptibility.
Finally, we have computed the nonlinear susceptibility, $\chi^{(3)}$,
defined from expansion of the magnetization in the direction of
the applied field: $M(H) = \chi^{(1)} H + \frac{1}{3!}\chi^{(3)} H^3
+ \cdots$. For UBe$_{13}$, in addition to linear splitting of the $5f^3$
magnetic doublet, a finite magnetic field induces a van Vleck splitting
and an overall shift of the $5f^2$ quadrupolar doublet, due to matrix
elements between the $\Gamma_3$ ground doublet and excited CEF
multiplets. We have included this effect by computing the $\Gamma_3$ energy
shifts perturbatively, up to fourth order in $H$. The first excited
CEF levels at $\Delta_{CEF} \approx 15{\rm mev}$~\cite{CEF} were assumed
to be degenerate $\Gamma_4$ and $\Gamma_5$ triplets, chosen as to allow the
van Vleck contribution to the linear susceptibility to quantitatively
match the UBe$_{13}$ data, as discussed in Ref.~\onlinecite{Cox87}.
This also fixed the position of the $\Gamma_1$ level. To make contact
with experiment, we fixed the ratio $T_K/\Delta_{CEF} \approx 0.055$,
which is necessary to account for the overestimated Kondo temperatures
in our calculations.

\begin{figure}
\centerline{
\vbox{\epsfxsize=80mm \epsfbox {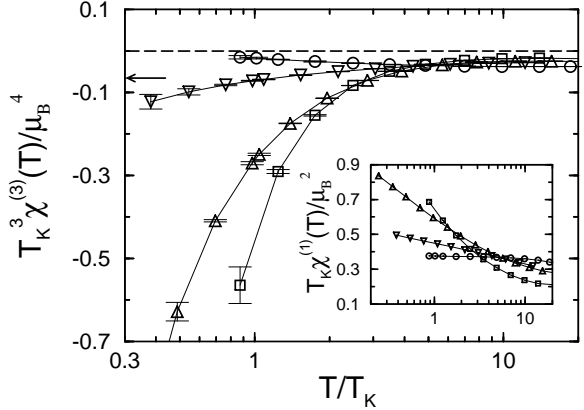}}
}\vspace{-5pt}
\caption{$\chi^{(3)}(T)$ along the (100) direction, for $\Gamma/D =
\pi/16$ and $\epsilon_f/D = -0.2\ (\circ), -0.08\ (\nabla), 0\ (\triangle)$,
and $0.2\ (\Box)$. The corresponding ground-state $5f^2$ weights
equal $0.71, 0.59, 0.5$, and $0.29$, respectively, while $T_K/D =
0.032, 0.02$, and $0.009$ for $|\epsilon_f|/D = 0, 0.08$, and $0.2$.
Inset: Linear susceptibility for the same set of parameters. For
$T_K \approx 10{\rm K}$, as appropriate for UBe$_{13}$,
$\chi^{(3)}(T=T_K) \approx -0.12 {\rm emu/mole T}^3$ for
$\epsilon_f/D = -0.08$, in close agreement with
experiment~\protect\cite{Ramirez94,Aliev95} (corresponding value for
UBe$_{13}$ is indicated by arrow). Similarly,
$\chi^{(1)}(0)\approx 0.014$emu/mole for the $\epsilon_f/D=-0.2$ case,
compared with an experimental value of $0.012\!-\!0.016$emu/mole.}
\end{figure}

%paragraph 17: Discussion of the calculated nonlinear susceptibility.
The resulting $\chi^{(3)}(T)$ curves are shown in Fig.~3. From these
calculations it is apparent that only for $\epsilon_f/D \approx -0.08$,
which corresponds to a ground-state $f^2$ occupancy of about 60\%, can
one obtain a curve quantitatively compatible with the pure UBe$_{13}$ data.
The curve for $\epsilon_f/D = -0.2$ (about 70\% ground-state $f^2$ weight)
resembles that for U$_{0.9}$Th$_{0.1}$Be$_{13}$. In contrast, the
$\epsilon_f/D = 0$ and $0.2$ curves have too strong a temperature
dependence to correspond with experiment, both for the nonlinear
{\em and} linear susceptibility (inset to Fig.~3). In every case, the
characteristic energy scale is too large for all of these mixed-valent
runs to account for the experimental data. We conclude that the scenario
of Aliev {\em et al.}~\cite{Aliev95}, suggesting that UBe$_{13}$ is more
strongly mixed valent than anticipated previously, and that doping with
Th increases the $5f^2$ weight (driving the system closer to the
quadrupolar Kondo limit), is qualitatively consistent with our results.

%paragraph 18: Inclusion of excited crystalline-electric-field multiplets.
The most severe omission from our model is a proper {\em dynamical}
treatment of excited CEF triplets (beyond perturbative shifting of the
bare $\Gamma_3$ levels). Although these triplets possess negligible
Boltzmann weights in the ionic limit, they can have significant
zero-temperature quantum weights for nonzero hybridization.
Due to their magnetic character, their inclusion can potentially render
the $\chi^{(3)}$ curves more negative even in the $5f^2$ limit. They
also introduce (even in the mixed-valent limit) a small energy scale
associated with crossover physics between ground-state screening of
the $\Gamma_3$ doublet, and collective screening of the entire Hund's
rule multiplet~\cite{Koga98}.

A.S. is grateful to H. Castella, M. Steiner, J. Wilkins, and Shiwei Zhang
for valuable discussions. Useful discussions with F. Aliev and M. Koga
are also acknowledged. This work was supported in part by a grant from the
US DOE, Office of Basic Energy Sciences, Division of Materials Research.
A.S. was supported in part by an OSU Postdoctoral Fellowship.


\begin{thebibliography}{9}
\bibitem{Maple95} For a review see, e.g., M.\ B.\ Maple {\em et al.},
   J.\ Low\ Temp.\ Phys. {\bf 99}, 223 (1995).
\bibitem{Cox87} D.\ L.\ Cox, Phys.\ Rev.\ Lett.\ {\bf 59}, 1240  (1987).
\bibitem{CZ97} For a detailed review, see
   D.\ L.\ Cox and F.\ Zawadowski, report no. 9704103 (to be published
   in Adv.\ in\ Phys.).
\bibitem{NB80} P.\ Nozi\'eres and A.\ Blandin, J.\ Phys.\ (Paris)
   {\bf 41}, 193 (1980).
\bibitem{Ramirez94} A.\ P.\ Ramirez {\em et. al.}, Phys.\ Rev.\ Lett.\
   {\bf 73}, 3018 (1994).
\bibitem{Aliev95} F.\ Aliev {\em et. al}, Europhys.\ Lett.\ {\bf 32},
   765 (1995).
\bibitem{Bethe_ansatz} N.\ Andrei and C.\ Destri, Phys.\ Rev.\ Lett.\
   {\bf 52}, 364 (1984); A.\ M.\ Tsvelik and P.\ B.\ Wiegman, Z.\ Phys.\ B
   {\bf 54}, 201 (1984); A.\ M.\ Tsvelik, J.\ Phys.\ C {\bf 18}, 159 (1985);
   P.\ D.\ Sacramento and P.\ Schlottmann, Phys.\ Rev.\ B {\bf 43},
   13294 (1991).
\bibitem{CFT} I.\ Affleck and A.\ W.\ W.\ Ludwig, Nucl.\ Phys.\ {\bf B360},
   641 (1991); A.\ W.\ W.\ Ludwig and I.\ Affleck, Phys.\ Rev.\ Lett.\
   {\bf 67}, 3160 (1991).
\bibitem{EK92} V.\ J.\ Emery and S.\ Kivelson, Phys.\ Rev.\ B {\bf 46},
   10812 (1992).
\bibitem{AY69} P.\ W.\ Anderson and G.\ Yuval, Phys.\ Rev.\ Lett.\ {\bf 23},
   89 (1969).
\bibitem{Haldane78} F.\ D.\ M.\ Haldane, J.\ Phys.\ C {\bf 11}, 5015 (1978).
\bibitem{SK93} Q.\ Si and G.\ Kotliar, Phys.\ Rev.\ Lett.\
   {\bf 70}, 3143 (1993); Phys.\ Rev.\ B {\bf 48}, 13881 (1993).
\bibitem{SS89} P.\ D.\ Sacramento and P.\ Schlottmann, Phys.\ Lett.\ A
   {\bf 142}, 245 (1989).
\bibitem{spherically_symmetric_ion} N.\ Read {\em et al.}, J.\ Phys.\ C
   {\bf 19}, 1597 (1986); A.\ C.\ Nunes {\em et al.}, J.\ Phys.\ C
   {\bf 19}, 1017 (1986).
\bibitem{CEF} S.\ M.\ Shapiro {\em et al.},
   J.\ Magn.\ Magn.\ Matt.\ {\bf 52}, 418 (1985);
   R.\ Felten {\em et al.}, Europhys.\ Lett.\ {\bf 2}, 323 (1986);
   S.\ L.\ Cooper {\em et al.}, Phys.\ Rev.\ B {\bf 35}, 2615 (1987).
\bibitem{Koga98} See Sec. 5.3.2 of Ref.~\onlinecite{CZ97}. The possibility
   of such a small crossover scale has been recently confirmed by M. Koga,
   by explicit numerical renormalization-group calculations.
\end{thebibliography}
\end{document}